\documentclass[a4paper]{jpconf}
\usepackage{graphicx}
\usepackage{bm}
\usepackage{epsfig}
\usepackage{graphics}
\usepackage{amsmath}
\usepackage{amssymb}
\usepackage{latexsym}
\usepackage{slashed,cite}
\usepackage{upgreek}
\usepackage[l]{floatflt}

\def\beq{\begin{equation}}
\def\eeq{\end{equation}}
\def\bea{\begin{eqnarray}}
\def\eea{\end{eqnarray}}

\newcommand\eqs[2]{Eqs.~(\ref{#1}) and (\ref{#2})}

\newcommand{\Eref}[1]{Eq.~(\ref{#1})}
\renewcommand{\Sref}[1]{Sec.~\ref{#1}}
\renewcommand{\Fref}[1]{Fig.~\ref{#1}}

\newcommand{\cref}[1]{Ref.~\cite{#1}}

\newcommand{\sEref}[2]{Eq.~(\ref{#1}{\small\sf {#2}})}

\newcommand{\beqs}{\begin{subequations}}
\newcommand{\eeqs}{\end{subequations}}

\def\lf{\left(}
\def\rg{\right)}
\renewcommand{\etal}{{\it et al.\/}}
\newcommand{\GeV}{{\mbox{\rm GeV}}}

\def\eec{\end{center}}
\def\bec{\begin{center}}
\newcommand{\eem}{\end{matrix}}
\newcommand{\bem}{\begin{matrix}}
\newcommand{\bfl}{\begin{flushleft}}
\newcommand{\efl}{\end{flushleft}}
\newcommand{\edm}{\end{displaymath}}
\newcommand{\bdm}{\begin{displaymath}}
\def\bea{\begin{eqnarray}}
\def\eea{\end{eqnarray}}

\newcommand{\ftn}{\footnotesize}

\newcommand{\ssz}{\scriptsize}

\newcommand{\Vhi}{\ensuremath{\widehat V_{\rm HI}}}

\def\lf{\left(}
\def\rg{\right)}

\def\llgm{\left\lgroup}
\def\rrgm{\right\rgroup}

\newcommand{\mP}{\ensuremath{m_{\rm P}}}
\newcommand{\Mgut}{\ensuremath{M_{\rm GUT}}}

\newcommand{\ld}{\ensuremath{\lambda}}

\newcommand{\kp}{\ensuremath{\kappa}}

\newcommand\vev[1]{\langle {#1} \rangle}

\newcommand{\Vjhi}{\ensuremath{V_{\rm HI}}}
\newcommand{\Hhi}{\ensuremath{\widehat H_{\rm HI}}}
\newcommand{\Khi}{\ensuremath{K}}

\newcommand{\lda}{\ensuremath{\lambda}}
\newcommand{\ldb}{\ensuremath{\lambda'}}

\renewcommand{\ns}{\ensuremath{n_{\rm s}}}
\newcommand{\as}{\ensuremath{\alpha_{\rm s}}}
\newcommand{\As}{\ensuremath{A_{\rm s}}}
\newcommand{\Ns}{\ensuremath{N_{\star}}}
\newcommand{\na}{\ensuremath{{N_a}}}

\newcommand{\rw}{\ensuremath{r_{0.002}}}

\newcommand{\ca}{\ensuremath{c_{\mathcal R}}}

\newcommand{\fr}{\ensuremath{f_\mathcal{R}}}

\newcommand{\fk}{\ensuremath{f_{\rm K}}}
\newcommand{\fp}{\ensuremath{f_{\rm P}}}

\newcommand{\ks}{\ensuremath{k_\star}}

\newcommand{\rcc}{\ensuremath{\mathcal{R}}}
\newcommand{\rce}{\ensuremath{\widehat{\mathcal{R}}}}
\newcommand{\Ve}{\ensuremath{\widehat{V}}}

\newcommand{\Ne}{\ensuremath{\widehat{N}}}

\newcommand{\msn}{\ensuremath{\what m_{\rm \dph}}}

\newcommand{\sg}{\ensuremath{\phi}}

\newcommand{\sgx}{\ensuremath{\phi_\star}}

\newcommand{\sgf}{\ensuremath{\phi_{\rm f}}}

\newcommand{\ldu}{\ensuremath{\uplambda}}

\newcommand{\se}{\ensuremath{\widehat\phi}}

\newcommand{\sex}{\ensuremath{\widehat{\phi}_\star}}
\newcommand{\sef}{\ensuremath{\widehat{\phi}_{\rm f}}}
\newcommand{\geu}{\ensuremath{\widehat g}}
\newcommand{\eph}{\ensuremath{\widehat \epsilon}}
\newcommand{\ith}{\ensuremath{\widehat \eta}}

\def\bbet{{\bar\beta}}
\def\al{{\alpha}}

\def\n{\bar{n}}

\def\Ka{K\"{a}hler potential}

\def\Km{K\"{a}hler manifold}
\def\Kaa{K\"{a}hler~}
\def\th{{\theta}}
\def\thb{{\bar\theta}}
\def\thn{{\theta_{\Phi}}}
\def\vth{{\vartheta}}

\newcommand{\phc}{\ensuremath{\Phi}}
\newcommand{\phcb}{\ensuremath{\bar\Phi}}

\def\thb{{\bar\theta}}
\def\thn{{\theta_{\Phi}}}

\newcommand{\kbba}{\ensuremath{{K_{221}}}}

\newcommand{\tkbba}{\ensuremath{{\widetilde K_{221}}}}
\newcommand{\tkba}{\ensuremath{{\widetilde K_{21}}}}

\newcommand{\kb}{\ensuremath{K_{2}}}
\newcommand{\kba}{\ensuremath{K_{21}}}

\newcommand{\mnfb}{\ensuremath{\mathcal{M}_{21}}}

\renewcommand{\tr}{{\mbox{\sf\ssz T}}}

\newcommand{\diag}{\mbox{\sf diag}}

\newcommand{\dph}{\ensuremath{\delta\phi}}
\newcommand{\what}{\ensuremath{\widehat}}
\newcommand{\wtilde}{\ensuremath{\widetilde}}

\newcommand{\Qef}{\ensuremath{\Lambda_{\rm UV}}}

\newcommand{\plk}{{\slshape\ssz Planck}}

\newcommand{\ma}{{MI}}
\newcommand{\mb}{{MII}}
\newcommand{\tmd}{{TM$_4$}}
\newcommand{\rs}{\ensuremath{\delta_\lambda}}

\newcommand{\cU}{\ensuremath{\mathcal{U}}}
\newcommand{\ch}{\ensuremath{\mathcal{P}}}
\newcommand{\bl}{\ensuremath{B-L}}
\newcommand{\mgut}{\ensuremath{M_{\rm GUT}}}
\newcommand{\mbl}{\ensuremath{M_{BL}}}
\newcommand{\Dex}{\ensuremath{\Delta_{\rm \star}}}

\newcommand\mtta[4]{\mbox{
$\llgm\bem #1 &#2 \cr #3& #4\eem\rrgm$}}
\newcommand\mttb[9]{\mbox{
$\llgm\bem #1 &#2&#3 \cr #4&#5&#6  \cr #7&#8&#9\eem\rrgm$}}

\newcommand\mtt[4]{\mbox{
$\llgm\bem #1 &#2 \cr #3& #4\eem\rrgm$}}

\begin{document}

\title{\boldmath $SU(2,1)/(SU(2)\times U(1))~B-L$ Higgs Inflation}

\author{C Pallis}

\address{Laboratory of Physics, Faculty of
Engineering, \\ Aristotle University of Thessaloniki,
Thessaloniki\\ GR-541 24, Greece}

\ead{kpallis@gen.auth.gr}

\date{\today}

\begin{abstract} We present a realization of Higgs inflation within Supergravity
which is largely tied to the existence of a pole of order two in
the kinetic term of the inflaton field. This pole arises due to
the selected Kaehler potential which parameterizes the
$SU(2,1)/(SU(2)\times U(1))$ manifold with scalar curvature ${\cal
R}_{21}=-6/N$. The associated superpotential includes, in addition
to the Higgs superfields, a stabilizer superfield, respects a
$B-L$ gauge and an $R$ symmetries and contains the first allowed
nonrenormalizable term. If the coefficient of this term is almost
equal to that of the others within about $10^{-5}$ and $N=2$, the
inflationary observables can be done compatible with the present
data. The tuning can be eluded if we modify the Kaehler potential
associated with the manifold above. In this case, inflation can be
realized with just renormalizable superpotential terms and results
to higher tensor-to-scalar ratios as $N$ approaches its maximum at
$N\simeq80$.

\end{abstract}

\section{Introduction}\label{sec:intro}

One of the major challenges in the are(n)a of inflationary model
building \cite{review} is the identification of the inflaton --
i.e., the scalar field causing inflation -- with one of the fields
already present in the fundamental theory. According to an
economical, predictive and highly attractive set-up, the inflaton
could play, at the end of its inflationary evolution, the role of
a Higgs field \cite{old, sm,  nmh, jhep, nmhk, univ, uvh, ighi} --
obviously we do not restrict our attention only to the electroweak
Higgs \cite{sm}. The relevant models (for a more complete list see
\cref{sor} and references therein) may be called \emph{Higgs
inflation} ({\sf HI}) for short. We find it technically convenient
to exemplify HI in our talk taking as reference theory an
``elementary" \emph{Grand Unified Theory} ({\sf GUT}) based on the
gauge group $G_{B-L}= G_{\rm SM}\times U(1)_{B-L}$ -- where
${G_{\rm SM}}$ is the gauge group of the \emph{Standard Model}
({\sf SM}), $B$ and $L$ denote baryon and lepton number
respectively. Despite its simplicity, $G_{B-L}$ is strongly
motivated by neutrino physics and leptogenesis mechanisms -- see
e.g. \cref{lept, univ}.

The spontaneous breaking of $U(1)_{B-L}$ requires the introduction
of a Higgs fields $\phc$ in a non-\emph{Supersymmetric}
({\sf\small SUSY}) framework or a conjugate pair of Higgs fields
$\phc$ and $\phcb$ in a SUSY context.  The Higgsflaton
\cite{kaloper} may be identified with the radial part of these
fields and is expected to obey a $\sg^4$ potential. Due to
inconsistency of that model with a number of observational and
theoretical requirements described in \Sref{obs}, we are obliged
to invoke some non-minimality. We below review -- in
Sec.~\ref{Fhi} -- the formulation of generalized non-minimal HI in
a non-SUSY framework. Then -- in Sec.~\ref{res0} -- we constrain
the parameters of two possible models, called E and T model in
\cref{prl}. In the next sections we investigate the
supersymmetrization of T model \cite{tmodel, pole} HI, following
the approach of \cref{sor}.

Throughout the text, the subscript $,\chi$ denotes derivation
\emph{with respect to} ({\sf w.r.t}) the field $\chi$, charge
conjugation is denoted by a star ($^*$) and we use units where the
reduced Planck scale $\mP = 2.43\cdot 10^{18}~\GeV$ is set equal
to unity.

\subsection{Formulating Non-SUSY Higgs Inflation}\label{Fhi}

The potential of the inflaton $\sg$ (i.e., the radial component of
$\phc$ assuming that the angular direction is somehow stabilized)
has to be of the well-known form
\beq V_{\rm
HI}=\ld^2(\sg^2-M^2)^2/16\simeq\ld^2\sg^4/16\>\>\>\mbox{for}\>\>\>M\ll\mP=1.\label{v4}\eeq
The corresponding action in the \emph{Jordan frame} ({\sf JF}),
takes the form:
\beqs\beq \label{action1} {\sf  S} = \int d^4 x
\sqrt{-\mathfrak{g}} \left(-\frac{\fr}{2}\rcc
+\frac{\fk}{2}g^{\mu\nu}
\partial_\mu \sg\partial_\nu \sg-
\Vjhi(\sg)+\cdots\right), \eeq
where $\mathfrak{g}$ is the determinant of the background
Friedmann-Robertson-Walker metric $g^{\mu\nu}$ with signature
$(+,-,-,-)$, $\rce$ is the space-time Ricci scalar,
$\vev{\fr}\simeq1$ to guarantee the ordinary Einstein gravity at
low energy and we allow for a kinetic mixing through the function
$\fk(\phi)$. The ellipsis includes the gauge interactions of $\sg$
which may be neglected at a tree level treatment. By performing a
conformal transformation \cite{old} according to which we define
the \emph{Einstein frame} ({\sf EF}) metric
$\geu_{\mu\nu}=\fr\,g_{\mu\nu}$ we can write ${\sf S}$ in the EF
as follows
\beq {\sf  S}= \int d^4 x
\sqrt{-\what{\mathfrak{g}}}\left(-\frac12
\rce+\frac12\geu^{\mu\nu} \partial_\mu \se\partial_\nu \se
-\Vhi(\se)+\cdots\right), \label{action} \eeq\eeqs
where hat is used to denote quantities defined in the EF. We also
introduce the EF potential, $\Vhi$, and canonically normalized
field, $\se$, derived as follows
\beq \label{VJe} \mbox{\sf (a)}\>\>\Vhi=
\frac{\Vjhi}{\fr^2}\>\>\>\mbox{and}\>\>\>\mbox{\sf
(b)}\>\>\frac{d\se}{d\sg}=J=\sqrt{\frac{\fk}{\fr}+{3\over2}\left({f_{
R,\sg}\over \fr}\right)^2}\,. \eeq
Naively, we expect that $\fr$ affects both $J$ and $\Vhi$ whereas
$\fk$ only $J$. Therefore, a suitable choice of $\fr$ \cite{old,
sm, nmh, nmi, prl} can lead to a plateau convenient for driving
HI. However, a plateau may also appear expressing $\Vjhi$ in terms
of $\se$ \cite{tmodel, pole} without the aid of $\fr$.

\subsection{Inflationary Observables -- Constraints} \label{obs}

The analysis of HI can be performed exclusively in the EF using
the standard slow-roll approximation as analyzed below, together
with the relevant observational and theoretical requirements that
should be imposed.

\subsubsection{} The number of e-foldings $\Ns$ that the scale
$\ks=0.05/{\rm Mpc}$ experiences during HI must be enough for the
resolution of the  problems of standard Big Bang, i.e.,
\cite{plcp}
\beqs\begin{equation} \label{Nhi}  \Ns=\int_{\sef}^{\sex}
d\se\frac{\Vhi}{\Ve_{\rm
HI,\se}}\simeq61.5+\frac14\ln\frac{\Vhi(\sex)^2}{g_{\rm
rh*}^{1/3}\Vhi(\sef)}\simeq55,\eeq
%=61.5+\frac14\ln\frac{\Vhi(\sex)^2}{g_{\rm rh*}^{1/3}\Vhi(\sef)}
% and horizon and flatness
where $g_{\rm rh*}=228.75$ and $\sex$ is the value of $\se$ when
$\ks$ crosses the inflationary horizon whereas $\se_{\rm f}$ is
the value of $\se$ at the end of HI, which can be found, in the
slow-roll approximation, from the condition
\beq{\sf max}\{\widehat\epsilon(\sg_{\rm
f}),|\widehat\eta(\sg_{\rm
f})|\}=1,\>\>\>~\mbox{where}\>\>\>\widehat\epsilon=
{1\over2}\left(\frac{\Ve_{\rm HI,\se}}{\Ve_{\rm
HI}}\right)^2\>\>\>\mbox{and}\>\>\>\widehat\eta= \frac{\Ve_{\rm
HI,\se\se}}{\Ve_{\rm HI}}\,.\label{sr}\eeq\eeqs

\subsubsection{} The amplitude $\As$ of the power spectrum of the curvature
perturbations generated by $\sg$ at  $k_{\star}$ has to be
consistent with data~\cite{plcp}, i.e.,
\begin{equation}  \label{Prob}
A_{\rm s}=\: \frac{1}{12\, \pi^2} \; \frac{\Ve_{\rm
HI}(\sex)^{3}}{\Ve_{\rm HI,\se}(\sex)^2} \simeq2.105\cdot
10^{-9}\,.
\end{equation}

\subsubsection{} The remaining inflationary observables (the spectral index $\ns$,
its running $\as$, and the tensor-to-scalar ratio $r$) have to be
consistent with the data \cite{plin,gws},  i.e.,
\begin{equation}  \label{nswmap}
\mbox{\sf
(a)}\>\>\ns=0.967\pm0.0074\>\>\>~\mbox{and}\>\>\>~\mbox{\sf
(b)}\>\>r\leq0.07,
\end{equation}
at 95$\%$ \emph{confidence level} ({\sf c.l.}) -- pertaining to
the $\Lambda$CDM$+r$ framework with $|\as|\ll0.01$. These
observables are estimated through the relations
\beq\label{ns} \mbox{\sf (a)}\>\>\ns=\: 1-6\eph_\star\ +\
2\ith_\star,\>\>\>\mbox{\sf (b)}\>\> \as
=\frac23\left(4\ith^2-(\ns-1)^2\right)-2\what\xi_\star\>\>\>~
\mbox{and}\>\>\>~\mbox{\sf (c)}\>\>r=16\eph_\star\,, \eeq
with $\widehat\xi={\Ve_{\rm HI,\se} \Ve_{\rm
HI,\se\se\se}/\Ve_{\rm HI}^2}$ -- the variables with subscript
$\star$ are evaluated at $\sg=\sgx$.

\subsubsection{} The effective theory describing HI has to remain
valid up to a UV cutoff scale $\Qef\simeq\mP$ to ensure the
stability of our inflationary solutions \cite{cutoff}, i.e.,
\beq \label{uv}\mbox{\sf (a)}\>\> \Vhi(\sgx)^{1/4}\leq\Qef
\>\>\>~\mbox{and}\>\>\>~\mbox{\sf (b)}\>\>\sgx\leq\Qef.\eeq

\subsection{Non-Minimal versus Pole-Induced HI}\label{res0}

The minimal model of HI defined by $\Vhi$ in \Eref{v4}, $\fk=1$
and $\fr=1$ is by now observationally excluded \cite{plin} since
it yields $\ns\simeq 0.947$ and $r\simeq0.28$ far away from the
restrictions in \Eref{nswmap}. Consequently, we have to invoke
some kind of non-minimality to reconcile HI based on $\Vhi$ in
\Eref{v4} with data. Actually, there are two available sources of
non-minimality as shown in \Eref{action1}. One due to $f_{\cal
R}\neq1$ and one due to $\fk(\phi)\neq1$. Below we review two
observationally viable models of HI which are based on successful
choices of either functions \cite{prl} -- for HI implemented with
a synergy between non-minimal $\fk$ and $\fr$ see \cref{ighi,univ,
nmhk}:

\subsubsection{(Standard) Non-Minimal or E-Model
HI} It can be realized \cite{sm, old, nmh}, if we adopt $\fk=1$
and $\fr=1+\ca\sg^2$. From \Eref{VJe} we obtain \beq V_{\rm
HI}\simeq\frac{\ld^2}{16\ca^2}\>\>\>\mbox{and}\>\>\>
J\simeq\frac{\sqrt{6}\ca\sg}{\fr}~~~\stackrel{\mbox{\ftn
(3b)}}{\Rightarrow}~~~\se=\sqrt{\frac{3}{2}}\ln\fr\>\>\>\mbox{or}\>\>\>\sg=\frac1{\sqrt{\ca}}\lf
e^{\sqrt{\frac{2}{3}}\widehat{\phi}}-1\rg^{1/2},\label{Je}\eeq
i.e., $\Vhi$ exhibits an almost flat plateau, whereas $\sg\sim
e^{\se}$ and thus, the name E-model HI. Indeed,
%=\frac{\ld^2}{16}\frac{\sg^4}{\fr^2}
from \eqs{Nhi}{sr} we find
\beqs\beq\eph\simeq {4/3\ca^2\sg^4},\>\>\>
\ith\simeq-{4/3\ca\sg^{2}}\>\>\>\mbox{and}\>\>\>\Ns\simeq{3\ca\sg_*^{2}/4}\,.\eeq
Therefore, $\sgf$ and $\sgx$ are found from the condition of
\Eref{sr} and the last equality above, as follows
\beq \label{sigs1}\sgf=(4/3\ca^2)^{1/4} \>\>\>\mbox{and}\>\>\>
\sgx=(4\Ns/3\ca)^{1/2}.\eeq\eeqs
Consequently, HI can be achieved even with subplanckian $\sg$
values for $\ca\gtrsim (4\Ne_*/3)^{1/2}$. Also the normalization
of \Eref{Prob} implies the following relation between $\ca$ and
$\ld$
\beq \label{Prob1} \ld\simeq{6}\sqrt{2\As}\pi\ca/{\Ns}\>
\Rightarrow\>
\ca\simeq2.25\cdot10^{4}\ld\>\>\>\mbox{for}\>\>\>\Ns\simeq55.\eeq
From \Eref{ns} we compute the remaining observables
\beq \label{nswmap1} \ns\simeq1-{2/\Ns}\simeq0.965, \>\>\> \as
\simeq{-2/\Ns^2}\simeq-6.4\cdot10^{-4}\>\>\> \mbox{and}\>\>\>
r\simeq{12/\Ns^2}\simeq4\cdot 10^{-3},\eeq
which are $\ca$-independent and in agreement with \Eref{nswmap}.

Although quite compelling, this model of HI is not consistent with
\Eref{uv} since $\Vhi^{1/4}\gg\Qef=\mP/\ca\ll\mP$ \cite{cutoff}.
This fact can be easily seen if we expand the second and third
term of ${\sf S}$ in the right-hand side of \Eref{action} about
$\vev{\phi}=0$ with results
\beq J^2 \dot\phi^2=\lf1-\ca\what{\sg}^2+6\ca^2\what{\sg}^2+
\ca^2\what{\sg}^{4}-\cdots\rg\dot\se^2~~\mbox{and}~~
\Vhi=\frac{\ld^2\what{\sg}^4}{2}\lf1-2\ca\what{\sg}^{2}+3\ca^2\what{\sg}^4-
\cdots\rg\cdot \label{Vexp}\eeq
%4\ca^3\what{\sg}^{6}+
Since the term which yields the smallest denominator for $\ca>1$
is  $6\ca^2\se^{2}$ we find $\Qef=\mP/\ca\ll\mP$ and so the
effective theory breaks down above it. For solutions proposed to
overcome this inconsistency rendering, mostly, the model less
predictive see \cref{uvh,ighi} and references therein.

\subsubsection{Pole-Induced or T-Model HI}\label{tmd}
In this case -- cf. \cref{tmodel, pole, sor} -- we introduce a
pole in $f_{\rm K}$, i.e., we set $\fr=1$ and $\fk=2N/\fp^2$ with
$\fp=1-\sg^2$ and $N>0$ -- possible coefficient of $\sg$ in $\fp$
can be trivially absorbed by a field redefinition. From \Eref{VJe}
we obtain
\beq \label{Jt} V_{\rm
HI}=\Vhi~~~\mbox{and}~~~J=\frac{\sqrt{2N}}{\fp}\>\>\stackrel{\mbox{\ftn
(3b)}}{\Rightarrow}\>\>
\se=\sqrt{\frac{N}{2}}\ln\frac{(1+\sg)}{(1-\sg)}\>\>\>\mbox{or}\>\>\>\sg=\tanh\frac{\se}{\sqrt{2N}},\eeq
i.e., we here obtain $\sg\sim\tanh\se$ and hence, the name
\emph{T-model} ({\sf \tmd}) HI. If we express $\Vjhi$ in terms of
$\se$, we easily infer that it experiences a stretching for
$\se>1$ which results to a plateau \cite{tmodel, sor}. Applying
\eqs{Nhi}{sr}, the slow-roll parameters and $\Ns$ read
\beqs\beq\label{nmci2b} \eph\simeq {16\fp^2/N\sg^2},\>\>\>
\ith\simeq{8\fp(3-5\sg^2)/N\sg^2}\>\>\>\mbox{and}\>\>\>
\Ns\simeq{N\sgx^2/4f_{\rm p\star}}.\eeq
Since $f_{\rm p\star}$ appears in the denominator, $\Ns$ increases
drastically as $\sgx$ approaches unity, assuring thereby the
achievement of efficient HI. Imposing the condition of \Eref{sr}
and solving then the latter equation w.r.t $\sgx$, we arrive at
\beq\label{nmci4b}\sgx=\sqrt{4\Ns}/\sqrt{4\Ns+N}\sim1\gg\sgf,\eeq\eeqs
i.e., HI is attained for $\sg<1$ thanks to the location of the
pole at $\sg=1$. On the other hand, \Eref{Prob} implies
\beq \label{Prob2} \ld\simeq{4\sqrt{6N\As}\pi/\Ns}\> \Rightarrow\>
\ld\sim10^{-5}\>\>\>\mbox{for}\>\>\>\Ns\simeq55.\eeq
Applying \Eref{ns} we find that the inflationary observables  can
be consistent with \Eref{nswmap}. Indeed,
\beq \label{ns2}\ns\simeq1-{2/\Ns}\simeq0.965, \>\>\> \as
\simeq-{2/\Ns^2}=9.5\cdot10^{-4} \>\>\> \mbox{and}\>\>\>
r\simeq{4N/\Ns^2}\leq0.07\>\>\>\Rightarrow\>\>\>N\leq80,\eeq
where the last bound is numerically obtained \cite{sor}.
Obviously, there is no problem with the effective theory here but
just a tuning related to the initial conditions -- see \Sref{res3}
below.

\paragraph{} \tmd\ is mostly analyzed assuming a gauge-singlet
inflaton -- cf. \cref{tmodel, pole}. We below describe its
possible realization by a (gauge non-singlet) Higgsflaton in the
context of \emph{Supergravity} ({\sf SUGRA}) -- see \Sref{sugra}
-- and then, in \Sref{res}, we analyze the inflationary behavior
of the proposed models. We conclude summarizing our results in
\Sref{con}.

\section{Supergravity Embeddings} \label{sugra}

\tmd\ $B-L$ HI can be supersymmetrized if we employ three chiral
superfields, a conjugate pair of Higgs superfields -- $z^1=\Phi$
and $z^2=\bar\Phi$ -- oppositely {charged} under the gauged
$U(1)_{B-L}$, and a gauge singlet $z^3=S$ which plays the role of
``stabilizer'' superfield.

In \Sref{sugra1} we present the basic formulation of a $B-L$ Higgs
theory within SUGRA and then we motivate in \Sref{Wsec} and
\ref{Ksec} the necessary ingredients for realizing $B-L$
pole-induced HI and describe in \Sref{Msec} the geometry of the
adopted moduli space.

\subsection{General Setup} \label{sugra1}

Denoting the scalar components of the various superfields by the
same superfield symbol, the EF action for $z^\al$'s  can be
written as
\beqs \beq\label{Saction1} {\sf S}=\int d^4x \sqrt{-\what{
\mathfrak{g}}}\lf-\frac{1}{2}\rce +K_{\al\bbet}\geu^{\mu\nu} D_\mu
z^\al D_\nu z^{*\bbet}-\Ve\rg, \eeq
where the kinetic mixing is controlled by the K\"ahler metric and
the covariant derivatives for scalar fields $z^\al$ defined
respectively as
\beq \label{Kab} K_{\al\bbet}={\Khi_{,z^\al
z^{*\bbet}}}>0\>\>\>\mbox{and}\>\>\>D_\mu z^\al=\partial_\mu
z^\al+ig A_{BL\mu} (B-L) z^\al,\eeq
where $g$ is the unified gauge coupling constant, $A_{BL\mu}$ is
the corresponding gauge field. On the other hand, the EF potential
$\what V$, depends on the K\"ahler potential $K$ and
superpotential, $W$, as shown from its explicit formula
\beq \Ve=\Ve_{\rm F}+ \Ve_{\rm D}\>\>\>\mbox{with}\>\>\> \Ve_{\rm
F}=e^{\Khi}\left(K^{\al\bbet}{\rm F}_\al {\rm F}^*_\bbet-3{\vert
W\vert^2}\right) \>\>\>\mbox{and}\>\>\>\Ve_{\rm D}= {1\over2}g^2
{\rm D}_{(B-L)}^2\cdot \label{Vsugra} \eeq
Here, a trivial gauge kinetic function is adopted. Also we use the
shorthand
\beq \label{Kinv} K^{\al\bbet}K_{\al\bar
\gamma}=\delta^\bbet_{\bar \gamma},\>\>{\rm F}_\al=W_{,z^\al}
+K_{,z^\al}W~~\mbox{and}~~{\rm D}_{B-L}=z_\al(B-L)
K^\al~~\mbox{with}~~ K^{\al}={\Khi_{,z^\al}}\cdot\eeq\eeqs
In this talk we concentrate on HI driven by $\Ve_{\rm F}$ along a
D-flat direction, and therefore we ignore any contribution from
$\Ve_{\rm D}$.

The interference of $K$ both in the $\phcb-\phc$ kinetic mixing
and in $\Ve_{\rm F}$ consists a complication w.r.t the non-SUSY
case since the presence of the pole in the kinetic terms affects
also $\Ve_{\rm F}$ -- cf. \cref{tkref}. This is rather problematic
and we show below how we arrange it in two ways.

\subsection{Superpotential}\label{Wsec}

Our proposal is based on the following superpotential\cite{jean}
\beq W=\lda S\lf \bar\Phi\Phi-M^2/2\rg/2-\ldb
S(\bar\Phi\Phi)^2,\label{Whi} \eeq
where $\lda, \ldb$ and $M$ are free parameters. $W$ is consistent
with a $U(1)_{B-L}$ and an $R$ symmetry acting on the various
superfields as follows
\beq (B-L)(S, \phcb, \Phi)=(0,-1,1)~~~\mbox{and}~~~R(S, \phcb,
\Phi)=(1,0,0).\eeq
$W$ leads to a $B-L$ phase transition since we expect that the
SUSY limit of $\Vhi$, after HI, for well behaved $K$'s, takes the
form
\beq V_{\rm
SUSY}\sim\frac14\ld^2\left|\phcb\phc-2(\phcb\phc)^2-{M^2}/2\right|^2
+\ld^2|S|^2\left|\phc(1-2\phcb\phc)+\phcb(1-2\phcb\phc)\right|^2+{\rm
D-terms},\eeq where we assume that $\ld\simeq\ldb$ -- see below.
The SUSY vacuum lies at the direction
\beq
\label{vev}\vev{S}=0~~\mbox{and}~~|\vev{\Phi}|=|\vev{\bar\Phi}|\simeq
M/\sqrt{2}~~\mbox{for}~~M\ll1,\eeq
%of the Superfields Under a Global $U(1)$
%
i.e., $U(1)_{B-L}$ is spontaneously broken via the vacuum
expectation values of $\Phi$ and $\bar\Phi$.

\subsection{K\"{a}hler Potentials}\label{Ksec}

Stabilization of the $S=0$ direction can be achieved without
invoking higher order terms, if we select:
\beq K_2=N_S\ln\lf1+|S|^2/N_S\rg~~~\mbox{with}~~~0<N_S<6,\eeq
which parameterizes the compact manifold $SU(2)/U(1)$ \cite{su11}.
On the other hand, to assure the presence of the pole within the
$\phcb-\phc$ system we select one of the following \Ka s
\beq \label{kba} \mbox{\sf
(a)}\>\>\kba=-N\ln\left(1-|\phc|^2-|\phcb|^2\right)\>\>\mbox{or}\>\>\>~\mbox{\sf
(b)}\>\> \tkba=\kba+K_{\rm H}+K_{\rm A},\eeq where the subscripts
``H" and ``A" stand for ``holomorphic" and ``antiholomorphic"
respectively and the relevant $K$'s are defined as
\beq  K_{\rm H}=(N/2)\ln(1-2\phcb\phc)\>\>\>\mbox{and}\>\>\>
K_{\rm A}=(N/2)\ln(1-2\phcb^*\phc^*).\eeq
Note that the present $N$ is twice that defined in \cref{sor}. We
observe that $\partial_{\al}\partial_{\bbet}\wtilde
K_{21}=\partial_{\al}\partial_{\bbet}K_{21}$ since
$\partial_{\bbet}K_{\rm H}=\partial_{\al}K_{\rm A}=0$ -- cf.
\cref{tkref}.
For both $K$'s, the D term due to $B-L$ symmetry is \beq{\rm
D}_{BL}= N\lf|\phc|^2-|\phcb|^2\rg/
\lf1-|\phc|^2-|\phcb|^2\rg=0 ~~\mbox{for}~~|\phc|=|\phcb|,\eeq
i.e., it can be eliminated during HI, if we identify the inflaton
with the radial parts of $\phc$ and $\phcb$.

The difference between $\kba$ and $\tkba$ arises from the factor
$e^{K}$ in \Eref{Vsugra}. Along the inflationary path,
$|\phc|=|\phcb|$, $\kba$ yields a denominator in $\Vhi$ which can
be almost cancelled out by tuning $\ldb/\lda$ in $W$ whereas
$\tkba$ does not lead to a denominator and so we can use $\ldb=0$
-- cf. \cref{tkref}. We examine the following working models:
\beq\bem \label{ms} \mbox{MI:}& \hspace*{-0.1in}
K=\kbba=\kb+\kba&\hspace*{-0.05in}\mbox{with}\>\>N=2&\hspace*{-0.05in}\mbox{and}\>\>\>W\>\>\mbox{with}\>\>\ldb\neq0;\\
\mbox{MII:}&\hspace*{-0.1in}
K=\tkbba=\kb+\tkba&\hspace*{-0.08in}\mbox{with~~free
$N$}&\hspace*{-0.06in}\mbox{and}\>\>W\>\>\mbox{with}\>\>\ldb=0.\eem\eeq

\subsection{Geometry of the \Ka s}\label{Msec}

For both $K=\kba$ and $\tkba$, we obtain the Bergmann metric,
i.e.,
\beq ds_{21}^2=K_{\al\bbet} dz^\al
dz^{*\bbet}=N\lf\frac{|d\phc|^2+|d\phcb|^2}{1-|\phc|^2-|\phc|^2}+
\frac{|\phc^*d\phc+\phcb^*d\phcb|^2}
{\lf1-|\phc|^2-|\phcb|^2\rg^{2}}\rg,\eeq
and the moduli-space scalar curvature
\beq{\cal
R}_{21}=-K^{\al\bbet}\partial_\al\partial_{\bbet}\ln\lf\det
M_{\phcb\phc}\rg=-{6}/{N},\eeq
where $M_{\phc\phcb}=(K_{\al\bbet})$ expresses the kinetic mixing
in the inflationary sector.

We can show that $ds_{21}^2$ parameterizes the manifold
$\mnfb=SU(2,1)/(SU(2)\times U(1))$. Indeed, an element $U$ of
$SU(2,1)$ satisfies the relations
\beq U^\dag \eta_{21} U=\eta_{21}\>\>\>\mbox{and}\>\>\>\det
U=1\>\>\>\mbox{with}\>\>\>\eta_{21}=\diag\lf1,1,-1\rg,\eeq and may
be written as -- cf. \cref{math} -- \beq\label{cUb} U=\cU \ch
\>\>\>\mbox{with} \>\>\>\cU=\mttb{1/\na}{0}{a}{\na b
a^*}{\na\gamma}{b}{\na \gamma a^*}{\na b^*}{\gamma} \>\>\>
\>\>\>\mbox{and}\>\>\>\ch=e^{i\vth}\mttb{d
}{f}{0}{-f^*}{d^*}{0}{0}{0}{e^{-3i\vth}},\eeq where
$\na=1/\sqrt{1+|a|^2}$ and the free parameters $a,b,\gamma,d$ and
$f$ are constrained as follows
\beq a,b,d,f\in \mathbb{C}, \gamma\in
\mathbb{R_+}\>\>\>\mbox{with}\>\>\>
|a|^2+|b|^2-\gamma^2=-1\>\>\>\mbox{and}\>\>\>|d|^2+|f|^2=1.\eeq
The former constraint ensures the hyperbolicity of $\mnfb$ whereas
the latter indicates the compactness of $SU(2)$ -- cf.
\cref{su11}. Obviously, $\ch\in SU(2)\times U(1)$ -- which is
subgroup of $SU(2,1)$ -- and depends on four (real) parameters.
Consequently, $\cU$ is a representative of $\mnfb$ depending on
four parameters.

The operation of $\cU$ on $\phc$ and $\phcb$ can be represented
via the isometric transformations
\beq ^\cU\phc=\frac{(1/\na)\phc+\na b^*a\phcb+\na a\gamma}{a^*
\phc +b^* \phcb+\gamma} \>\>\>\mbox{and}\>\>\> ^\cU\phcb=\frac{\na
\gamma \phcb+\na b}{a^* \phc +b^* \phcb+\gamma}\,,
\label{tr21}\eeq where the $\phc$ and $\phcb$ independent
parameters originate from the lines of $\cU^\dag$ and have the
following $\bl$ charges
\beq \label{blb} (\bl)(a, b,\gamma)=(1,-1,0). \eeq
It is straightforward to show that $^\cU ds_{21}^2=ds_{21}^2$ and
so, $\kba$ and $\tkba$ parameterize $\mnfb$. Moreover, $\kba$ in
remains invariant under \Eref{tr21}, up to a \Kaa transformation,
i.e.,
\beq ^\cU \kba=\kba+\Lambda+\Lambda^*\>\>\>\mbox{and}\>\>\>^\cU W=
We^{-\Lambda}\,\>\>\mbox{with}\>\>\Lambda=N\ln(a^* \phc +b^*
\phcb+\gamma),\eeq
whereas $\tkba$ does not enjoy such an invariance. Note in passing
that  the employed $K$'s for E-model HI do not parameterize any
specific manifold and so those models are not so predictive -- cf.
\cref{nmh,univ,uvh,jhep,nmhk}.

\section{Inflation Analysis}\label{res}

We below, in \Sref{res1}, specify the form of the inflationary
potential derived by the SUGRA settings above and then, we
investigate its stability -- see \Sref{res2} -- and derive its
observational outputs -- see \Sref{res3}.

\subsection{Inflationary Potential}\label{res1}

If we use the parameterizations  \beq \Phi=\phi
e^{i\theta}\cos\theta_\Phi\>\>\> \mbox{and}\>\>\>\bar\Phi=\phi
e^{i\thb}\sin\theta_\Phi\>\>\>
\mbox{with}\>\>\>0\leq\thn\leq{\pi}/{2}~~~\mbox{and}~~~S= \lf{s
+i\bar s}\rg/{\sqrt{2}},\eeq
we can easily verify that a D-flat direction is \beq
\theta=\thb=0,\>\thn={\pi/4}\>\>\>\mbox{and}\>\>\>S=0,\label{inftr}\eeq
which can be qualified as inflationary path. The only surviving
term of $\Ve_{\rm F}$ along it is
\beq \Vhi= e^{K}K^{SS^*}\, |W_{,S}|^2=
\frac{\lda^2}{16}\cdot\left.\begin{cases}
{\lf\sg^2-(1+\rs)\sg^4-M^2\rg^2}/{\fp^{N}},\\
\lf\sg^2-M^2\rg^2.\end{cases}\right.\eeq
%\}
We can easily verify that \ma\ for $N=1$ and $M=\rs=0$ and \mb\
for $M=0$ share the same expression for $\Vhi$, given in
\Eref{v4}, which is actually the potential adopted for \tmd\
\cite{tmodel}.

To obtain \tmd, though, we have to establish the correct
non-minimal kinetic mixing shown in \Eref{Jt}. To this end, we
compute $K_{\al\bbet}$ along the path in \Eref{inftr} which takes
the form
\beq \lf K_{\al\bbet}\rg=\diag\lf
M_{\phc\phcb},K_{SS^*}\rg~~\mbox{with}~~
M_{\phc\phcb}=\frac{\kp\sg^2}{2}\mtta{2/\sg^2-1}{1}{1}{2/\sg^2-1},
\>\> \kp=\frac{N}{\fp^{2}}\eeq
and $K_{SS^*}=1$. Then we diagonalize $M_{\phc\phcb}$ via a
similarity transformation as follows:
\beq U_{\phc\phcb} M_{\phc\phcb} U_{\phc\phcb}^\tr =\diag\lf
\kp_+,\kp_-\rg,\>\>\>\mbox{where}\>\>\>U_{\phc\phcb}=
\frac{1}{\sqrt{2}}\mtt{1}{1}{-1}{1},\>\>
\kp_+=\kp\>\>\>\mbox{and}\>\>\> \kp_-=\kp\fp\,. \eeq
Canonically normalizing the various fields, we obtain the desired
form of $J$ in \Eref{Jt} whereas for the remaining fields we get
\beq \widehat{\theta}_+
=\sqrt{\kp_+}\sg\theta_+,\>\>\widehat{\theta}_-
=\sqrt{{\kp_-}}\sg\theta_-,\>\>\widehat \theta_\Phi =
\sg\sqrt{2\kp_-}\lf\theta_\Phi-{\pi}/{4}\rg
\>\>\>\mbox{and}\>\>\>\lf\what s,\what{\bar s}\rg=\lf s,\bar
s\rg,\eeq
%%J=\sqrt{2\kp_+}\>\>\Rightarrow\>\>\sg=\tanh\frac{\se}{2\sqrt{N}},\>\>
where $\th_{\pm}=\lf\bar\th\pm\th\rg/\sqrt{2}$.

\subsection{Stability of Inflationary Direction}\label{res2}

\renewcommand{\arraystretch}{1.3}

\begin{table}[!t]
\caption{\sl Mass-squared spectrum along the inflationary
trajectory}
\begin{center}
\lineup{\small
\begin{tabular}{c@{\hspace{0.1cm}}c@{\hspace{0.3cm}}c@{\hspace{0.3cm}}c@{\hspace{0.3cm}}c}
\br {\sc Fields}&{\sc Eigen-}& \multicolumn{3}{c}{\sc Masses
Squared}\\\cline{3-5} &{\sc states}& &
\multicolumn{1}{c}{$K=\kbba$}&\multicolumn{1}{c}{$K=\tkbba$}\cr
\mr
%\hspace*{2.cm}
%
2 real&$\widehat\theta_{+}$&$m_{\widehat\theta+}^2$&
\multicolumn{2}{c}{$3\Hhi^2$}\cr
scalars&$\widehat \theta_\Phi$ &$\widehat m_{
\theta_\Phi}^2$&\multicolumn{2}{c}{$M^2_{BL}+6\Hhi^2(1+2/N-2/N\sg^2)$}
\cr
1 complex&$s, {\bar{s}}$ &$ \widehat m_{
s}^2$&\multicolumn{1}{c}{$6\Hhi^2(1/N_S-8(1-\sg^2)/N+N\sg^2/2$}&\multicolumn{1}{c}{$6\Hhi^2(1/N_S-4/N$}\cr
scalar&&&\multicolumn{1}{c}{$+2(1-2\sg^2)+4\sg^2/N)$}&\multicolumn{1}{c}{$+2/N\sg^2+2\sg^2/N)$}\cr\mr
1 gauge boson &{$A_{BL}$}&{$M_{BL}^2$}&
\multicolumn{2}{c}{$2Ng^2\sg^2/\fp^2$}\cr\mr
$4$ Weyl  & $\what \psi_\pm$ & $\what m^2_{ \psi\pm}$&
\multicolumn{2}{c}{${12\fp^2\Hhi^2/N^2\sg^2}$}\cr
spinors&$\ldu_{BL},
\widehat\psi_{\Phi-}$&$M_{BL}^2$&\multicolumn{2}{c}{$2Ng^2\sg^2/\fp^2$}\cr
\br
\end{tabular}}
\end{center}
\label{spectrum}
\end{table}

\renewcommand{\arraystretch}{1.}

To consolidate our inflationary setting we have to check the
stability of the trajectory in \Eref{inftr} w.r.t the fluctuations
of the non-inflaton fields. Approximate, quite precise though,
expressions for the relevant mass-squared spectrum are arranged in
Table~\ref{spectrum} and assist us to appreciate the role of
$0<N_S<6$ in retaining positive and heavy enough $\what m^2_{s}$
for both $K$'s. In Table~\ref{spectrum} we display also the
masses, $M_{BL}$, of the gauge boson $A_{BL}$ -- which signals the
fact that $U(1)_{B-L}$ is broken during HI  and so, no topological
defects are produced during the subsequent $B-L$ phase transition
-- and the masses of the corresponding fermions. We determine $M$
demanding that the GUT scale $\Mgut\simeq2/2.433\times10^{-2}$ is
identified with $M_{BL}$ at the vacuum -- in \Eref{vev}, I.e.,
\beq \label{Mg} \vev{M_{BL}}={\sqrt{2N}gM/
\vev{\fp}}=\mgut\>\>\Rightarrow\>\>M\simeq{\mgut}/{g\sqrt{2N}}
\>\>\mbox{with}\>\>\>g\simeq0.7.\eeq
It can be verified, finally, that the one-loop radiative
corrections \`{a} la Coleman-Weinberg to $\Vjhi$ can be kept under
control -- see \cref{jhep}.

\subsection{Inflationary Observables - Results}\label{res3}

For MII we obtain the same results as in the non-SUSY case
analyzed in \Sref{tmd}. For MI approximate results are presented
in \cref{sor}. From the presented expressions there we infer that
$\Ns$ is well approximated by \Eref{nmci2b} by setting $N=2$
whereas the slow-roll parameters and the observables display a
dependence on $\rs$ and $M$ which modifies drastically the
predictions of the model compared to those of \tmd. MI resembles
better the behavior of the model studied in \cref{eno5} realized
by a gauge singlet, though.

We here discuss our numerical results. Let us initially recall
that our inflationary scenaria depend on the parameters:
$$M,\>\lda\>\>\mbox{and}\>\>\rs\>\>\mbox{for MI,
or}\>\>N\>\>\mbox{for MII}.$$ $M$ can be uniquely determined
imposing \Eref{Mg} whereas $\lda$ together with $\sgx$ can be
found enforcing \eqs{Nhi}{Prob}. By varying the one remaining
parameter for each model we obtain the allowed curves for MI and
MII (dashed and solid lines respectively) in the $\ns-r_{0.002}$
plane as designed in \Fref{fig1}. These outputs are compared with
the observationally favored corridors at $68\%$ [$95\%$] c.l.
depicted by the dark [light] shaded contours \cite{plin}. Some
parameters of both models for one representative value of the free
parameter are also arranged in the Table of \Fref{fig1}.

\begin{figure}[t]\vspace*{-0.6cm}
\begin{minipage}{75mm}
\includegraphics[height=9cm,angle=-90]{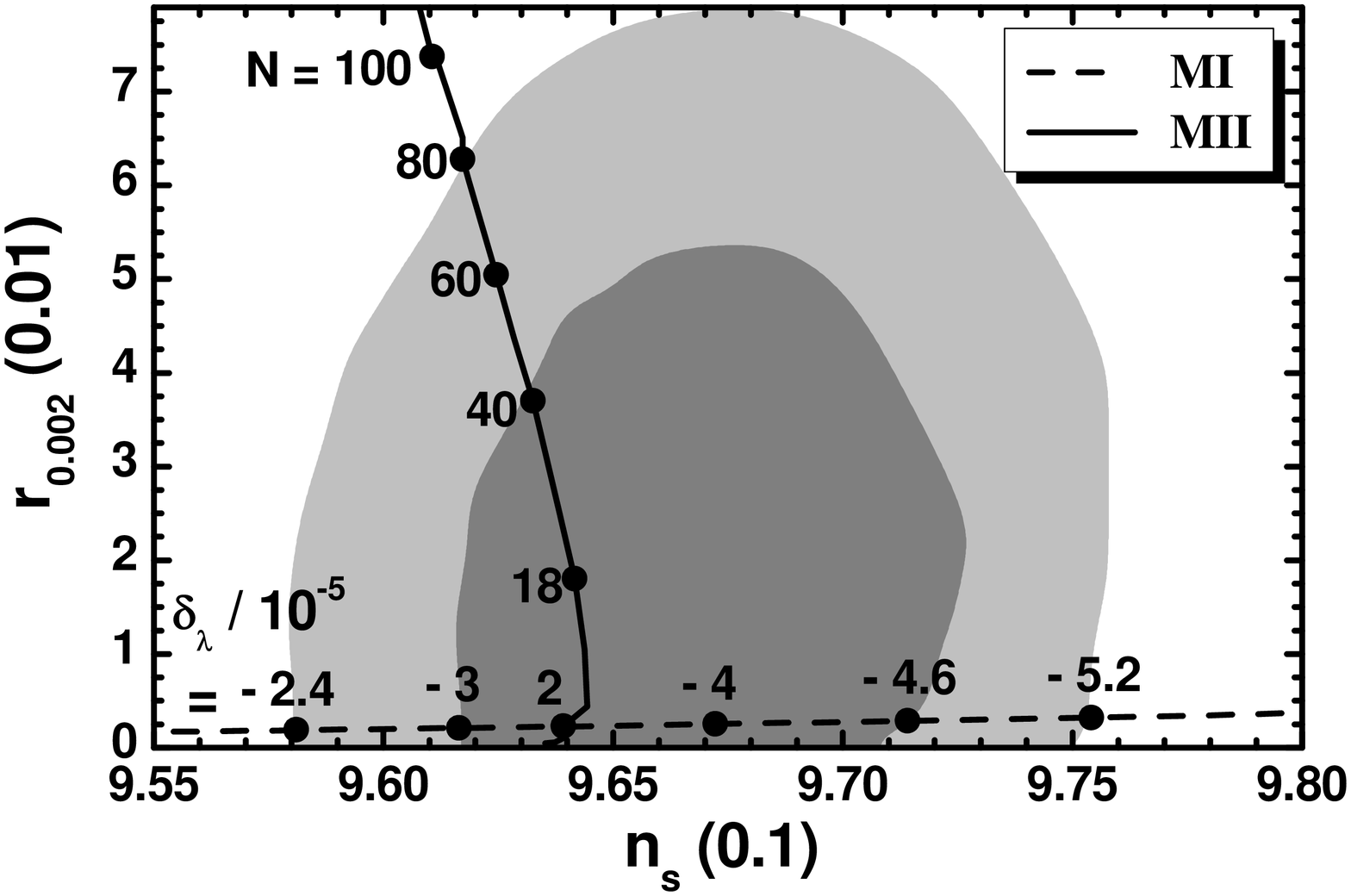}
\end{minipage}
\hfil
\begin{minipage}{65mm}
\begin{center}
\lineup{\small
\begin{tabular}{@{}ccc@{}}
\br {\sc Model:} &{\ma}&{\mb}\cr\mr
${\rs/10^{-5}}$&$-4$&{$-$}\cr $N$&{$2$}&$18$\cr\mr
$\sgx/0.1$&$9.9564$&$9.6322$\cr
$\Dex (\%)$&$0.44$&$3.7$\cr
$\sgf/0.1$&{$7.1$}&$4.3$\cr\mr
$\ld/10^{-5}$&$3.8$&$10.7$\cr
$M/10^{-3}$&{$5.87$}&$1.96$\cr\mr
$\ns/0.1$&$9.67$&$9.64$\cr
$-\as/10^{-4}$&$7.1$&$6.4$\cr
$r/10^{-2}$&$0.28$&$2$\cr \br
\end{tabular}}
\end{center}
\end{minipage}
\caption{\sl Allowed curves in the $\ns-\rw$ plane fixing
$\mbl=\mgut$ for \ma\ and various $\rs$'s indicated on the dashed
line or \mb\ and various $N$'s indicated on the solid line. The
marginalized joint $68\%$ [$95\%$] c.l. regions from \plk\ TT, TE,
EE+lowE+lensing, {\sffamily BK14} and BAO data \cite{plin} are
depicted by the dark [light] shaded contours. The relevant field
values, parameters and observables corresponding to
$\rs=-4\cdot10^{-5}$ for MI and $N=18$ for MII are listed in the
Table.} \label{fig1}\end{figure}

As already anticipated in both models $\sgx\sim 1$ and so a tuning
emerges which can be qualified by computing $\Dex={1 -\sgx}$. For
MI $\Dex$ is close to $0.4\%$ and the whole observationally
favored range can be covered for $\rs$'s close to $10^{-5}$
whereas $r$ remaining below $0.01$. I.e.,
\beq 2.4\lesssim\frac{-\rs}{10^{-5}}\lesssim
5.2,\>\>\>4.6\gtrsim\frac{\Dex}{10^{-3}}\gtrsim
4.1,\>\>\>5.4\lesssim\frac{-\as}{10^{-4}}\lesssim
8.6\>\>\>\mbox{and}\>\>\> 2.1\lesssim
\frac{r}{10^{-3}}\lesssim3.4\,.\eeq
Fixing $\ns=0.967$ we find $\rs=10^{-5}$ and $r=0.0028$. For MII
$\ns$ is concentrated a little lower than its central value and
$r$ increases with $N$ and $\Dex$. I.e,
\beq 0.962\lesssim\ns\lesssim0.964,\>\>\>0.5\lesssim N\lesssim
80,\>\>\>0.45\lesssim{\Dex}/{10^{-2}}\lesssim
13.6\>\>\>\mbox{and}\>\>\> 0.0025\lesssim {r}\lesssim0.07\,.\eeq
We find a robust upper bound on $N$, $N\lesssim80$, derived by the
one in \sEref{ns}{c}.

The mass of the inflaton at the SUSY vacuum of \Eref{vev} is given
by
\beq \msn=\left\langle\Ve_{\rm HI,\se\se}\right\rangle^{1/2}=
\left\langle \Ve_{\rm
HI,\sg\sg}/J^2\right\rangle^{1/2}\simeq\frac{\lda
M}{2\sqrt{N}}\cdot
\begin{cases}
\sqrt{1-3M^2}\simeq(1.7-2.1)\cdot10^{11}~\GeV&\mbox{for \ma,}\\
\vev{\fp}~~~~~\simeq(1.8-1.9)\cdot10^{11}~\GeV&\mbox{for
\mb.}\end{cases}\eeq
These ranges let open the possibility of non-thermal leptogenesis
\cite{lept}, if we introduce a suitable coupling between the
$\bar\Phi$ and the right-handed neutrinos $N^c_i$, $\bar\Phi
N_i^{c2}$ -- cf. \cref{univ, uvh, ighi}.

\section{Conclusions} \label{con}

We presented two models (MI and MII) of HI within SUGRA employing
\Ka s which parameterize the $SU(2,1)/(SU(2)\times U(1))$ \Km. The
Higgs fields implement the breaking of a gauge $U(1)_{B-L}$
symmetry at a scale $M$ which may assume a value compatible with
the MSSM unification. Both models display a kinetic mixing with a
pole of order two in the inflaton sector and respect an $R$ and
the gauge symmetries. Within \ma\ we employ $K=\kbba$ in
\sEref{kba}{a} and the first allowed non-renormalizable term in
$W$ of \Eref{Whi}. Any observationally acceptable $\ns$ is
attainable by tuning $\rs\simeq10^{-5}$, whereas $r\sim 10^{-3}$.
Within MII we use $K=\tkbba$ in \sEref{kba}{b} and only
renormalizable terms in $W$ of \Eref{Whi}. We find
$\ns\simeq0.961-0.963$ and $r$ increases with $N$, which is
related to ${\cal R}_{21}=-6/N$, and is bounded from above,
$N\leq80$. The inflaton mass is collectively confined into the
range $(1.7-2.1)\cdot10^{11}~\GeV$.

\ack Work supported by the Hellenic Foundation for Research and
Innovation (H.F.R.I.) under the ``First Call for H.F.R.I. Research
Projects to support Faculty members and Researchers and the
procurement of high-cost research equipment grant'' (Project
Number: 2251).

\section*{References}

\def\ijmp#1#2#3{{\sl Int. Jour. Mod. Phys.}
{\bf #1},~#3~(#2)}
\def\plb#1#2#3{{\sl Phys. Lett. B }{\bf #1}, #3 (#2)}
\def\prl#1#2#3{{\sl Phys. Rev. Lett.}
{\bf #1},~#3~(#2)}
\def\rmp#1#2#3{{Rev. Mod. Phys.}
{\bf #1},~#3~(#2)}
\def\prep#1#2#3{{\sl Phys. Rep. }{\bf #1}, #3 (#2)}
\def\prd#1#2#3{{\sl Phys. Rev. D }{\bf #1}, #3 (#2)}
\def\prdn#1#2#3#4{{\sl Phys. Rev. D }{\bf #1}, no. #4, #3 (#2)}
\def\prln#1#2#3#4{{\sl Phys. Rev. Lett. }{\bf #1}, no. #4, #3 (#2)}
\def\npb#1#2#3{{\sl Nucl. Phys. }{\bf B#1}, #3 (#2)}
\def\npps#1#2#3{{Nucl. Phys. B (Proc. Sup.)}
{\bf #1},~#3~(#2)}
\def\mpl#1#2#3{{Mod. Phys. Lett.}
{\bf #1},~#3~(#2)}
\def\jetp#1#2#3{{JETP Lett. }{\bf #1}, #3 (#2)}
\def\app#1#2#3{{Acta Phys. Polon.}
{\bf #1},~#3~(#2)}
\def\ptp#1#2#3{{Prog. Theor. Phys.}
{\bf #1},~#3~(#2)}
\def\n#1#2#3{{Nature }{\bf #1},~#3~(#2)}
\def\apj#1#2#3{{Astrophys. J.}
{\bf #1},~#3~(#2)}
\def\mnras#1#2#3{{MNRAS }{\bf #1},~#3~(#2)}
\def\grg#1#2#3{{Gen. Rel. Grav.}
{\bf #1},~#3~(#2)}
\def\s#1#2#3{{Science }{\bf #1},~#3~(#2)}
\def\ibid#1#2#3{{\it ibid. }{\bf #1},~#3~(#2)}
\def\cpc#1#2#3{{Comput. Phys. Commun.}
{\bf #1},~#3~(#2)}
\def\astp#1#2#3{{Astropart. Phys.}
{\bf #1},~#3~(#2)}
\def\epjc#1#2#3{{Eur. Phys. J. C}
{\bf #1},~#3~(#2)}
\def\jhep#1#2#3{{\sl J. High Energy Phys.}
{\bf #1}, #3 (#2)}
\newcommand\jcap[3]{{\sl J.\ Cosmol.\ Astropart.\ Phys.\ }{\bf #1}, #3 (#2)}
\newcommand\jcapn[4]{{\sl J.\ Cosmol.\ Astropart.\ Phys.\ }{\bf #1}, no. #4, #3 (#2)}
\newcommand{\arxiv}[1]{\texttt{arXiv:#1}}
\newcommand{\astroph}[1]{{\tt astro-ph/#1}}
\newcommand{\hepph}[1]{{\tt hep-ph/#1}}

\end{document}